\documentstyle[preprint,eqsecnum,aps,epsf,psfig]{revtex}
\newif\iftightenlines\tightenlinesfalse
\tightenlines\tightenlinestrue
\begin{document}

\title{Lepton Fluxes from Atmospheric Charm}
\author{L. Pasquali$^1$, M. H. Reno$^{1}$ and I. Sarcevic$^2$}
\address{
$^1$Department of Physics and Astronomy, University of Iowa, Iowa City,
Iowa 52242\\
$^2$Department of Physics, University of Arizona, Tucson, Arizona
85721}

\maketitle

\begin{abstract}

We re-examine the charm contribution to atmospheric lepton fluxes in the
context of perturbative QCD.  We include next-to-leading order
corrections and discuss theoretical uncertainties due to the
extrapolations of the gluon distributions at small-$x$.  
We show that the charm contribution to the atmospheric muon flux becomes
dominant over the conventional contribution from $\pi$ and $K$ decays
at the energies of about $10^5$ GeV.  
We compare our fluxes with previous calculations.

\end{abstract}

\section{INTRODUCTION}

Neutrino and muon fluxes from cosmic ray interactions with the Earth's 
atmosphere
have been  topics of considerable experimental and theoretical
interest \cite{review}. At energies near 1 GeV, the IMB \cite{imb}, 
Kamiokande 
\cite{kam} and Soudan \cite{soudan} experiments
detect an excess of $\nu_e$ relative to $\nu_\mu$ in the atmospheric 
neutrinos.  Recent results from SuperKamiokande \cite{superk} appear to 
confirm 
this observation.  
At these energies, 
leptonic decays of charged pions and leptonic and semileptonic decays
of kaons are responsible for the lepton fluxes, the so-called 
``conventional'' lepton flux.
Currently, it is believed that 
the conventional flux dominates until energies
of about $10^3$ TeV, when the effects of 
atmospheric charm production
and decay become important contributions to the lepton fluxes. 
The issue of where the charm contributions dominate is of interest,
in part, because this is an energy regime accessible to large
underground experiments \cite{ghs}.  
Recent results from Fr\'ejus \cite{rhode}, Baksan
\cite{kgf} and other experiments \cite{other} show an excess relative to
the conventional muon flux in the 10
TeV energy range. This may be an indication of 
a charm 
contribution at lower energies that expected.  
One of the main goals of the 
neutrino experiments such as AMANDA \cite{amanda}, Antares \cite{antares}, 
Nestor \cite{nestor} and at 
Lake Baikal \cite{baikal}
are searches for muon neutrinos from extragalactic
neutrino sources for which atmospheric 
neutrinos and muons  present the main background.   

Lepton fluxes from atmospheric charm have been calculated previously
\cite{tig,volkova,vfgs,bugaev} 
for specific models of charm particle production.
Here, we calculate the leptonic flux from charm in the context of
perturbative QCD.
We include next-to-leading order radiative corrections and we study 
the importance of small-$x$ behavior of the parton distribution functions.  
We emphasize
the uncertainties inherent in the necessary extrapolation of cross sections
and energy distributions beyond the experimentally measured regime.
We use the comparison with low-energy charm production 
data to constraint some of the 
theoretical uncertainties, such as the charm quark mass and the
factorization and renormalization scale dependence.  
We compare 
our results to the earlier work
on the prompt muons from charm including a recent calculation \cite{tig}
calculated using PYTHIA Monte Carlo program \cite{pythia}.

In the next section, we describe the framework for the
calculation of the lepton fluxes. In Section III, we
focus on the charmed quark contribution.
In Section IV, we present our results for the fluxes and 
compare with other calculations. We conclude in Section V.

\section{Lepton Flux Calculation}

Particle fluxes are determined by solving the 
coupled differential equations
that account for production, decays and interactions of the particles.
The general form of the cascade equations describing the propagation of
particle $j$ through column depth $X$ is given by\cite{book,lipari}
\begin{equation}
{{\rm d}\phi_j\over {\rm d}X} = -{\phi_j\over \lambda_j}
-{\phi_j\over \lambda^{(dec)}_j}+\sum_k S(k\rightarrow j)
\end{equation}
where $\lambda_j$ is the interaction length, $\lambda_j^{(dec)}\simeq
\gamma c \tau_j \rho(X)$ is the
decay length, accounting for time dilation factor $\gamma$ 
and expressed in terms
of g/cm$^2$ units. The density of the atmosphere is $\rho(X)$ and 
\begin{equation}
S(k\rightarrow j)= \int_E^\infty {\rm d}\,E_k
{\phi_k(E_k,X)\over \lambda_k(E_k)}
 {{\rm d}n_{k\rightarrow j}(E;E_k)
\over {\rm d}E}\ . 
\end{equation}
In Eq. (2.2), d$n$/d$E$ refers to
either the production distribution $1/\sigma_{k}\cdot
$d$\sigma_{k\rightarrow j}$
/d$E$
or decay distribution  $1/\Gamma_k\cdot$d$\Gamma_{k\rightarrow j}$/d$E$
(where $\lambda_k\rightarrow \lambda_k^{(dec)}$ in Eq. (2.2))
as a function of the
energy $E$ of the outgoing particle $j$. 

It is possible to solve these equations numerically, however, it has
been shown \cite{tig} that the same results can be 
obtained with an  analytic solution which 
was derived by noticing that 
the energy dependence of the fluxes approximately factorizes from the
$X$ dependence. Consequently, one can rewrite
\begin{eqnarray}
 S(k\rightarrow j)& \simeq &{\phi_k(E,X)\over \lambda_k(E)}
\int_E^\infty {\rm d}\,E_k
{\phi_k(E_k,0)\over\phi_k(E,0)}{\lambda_k(E)\over \lambda_k(E_k)}
 {{\rm d}n_{k\rightarrow j}(E;E_k)
\over {\rm d}E}\\ \nonumber
&\equiv &  { \phi_k(E,X)\over \lambda_k(E)} Z_{kj}(E)\ .
\end{eqnarray}
It is often convenient to write $Z_{kj}$ in terms of an integral
over $x_E\equiv E/E_k$, so
\begin{equation}
Z_{kj}(E) = \int_0^1 {{\rm d}x_E\over x_E} {\phi_k(E/x_E,0)\over \phi_k(E,0)}
{\lambda_k(E)\over \lambda_k(E/x_E)}{{\rm d}n_{k\rightarrow j}(E/x_E)
\over {\rm d}x_E}
\ .
\end{equation}

In the limits where the flux has a single power law energy behavior,
the interaction lengths are energy independent and
the differential distribution is scaling (energy independent),
the $Z$-moment $Z_{kj}(E)$ is independent of energy.
In practice, the $Z$-moments have a weak energy dependence
because d$n$/d$x_E$ depends on $E_k$, the interaction lengths
$\lambda$ are not energy independent, and in general, $\phi_k(E)$ is not
a constant power law in energy over the full energy range.
The cosmic ray flux can be represented by the following flux of
primary nucleons at $X=0$:
\begin{eqnarray}
\phi_p(E,X=0) [{\rm cm^{-2}s^{-1}sr^{-1}GeV^{-1}}]\  =\ &
1.7\ (E/{\rm GeV})^{-2.7}\quad E<E_0\\ \nonumber
& 174\ (E/{\rm GeV})^{-3}\quad E\ge E_0\ ,
\end{eqnarray}
where
$E_0=5\cdot 10^6$ GeV \cite{pal,jacee}.
At these energies, we 
assume isotropy of the flux \cite{clay}.

The detailed solutions to the cascade equations can be found, for
example, in Refs. \cite{book} and \cite{lipari}.
Following Ref. \cite{tig}, we assume that the incident cosmic ray
flux can be represented by protons.
The flux results, in high energy and low energy regimes for lepton flavor
$\ell=\nu_\mu,\ \nu_e$ or
$\mu$ due to proton production of hadron $j$ followed by
$j$ decay into  $\ell$ are
\begin{equation}
\phi_\ell^{j,high} =  
{ Z_{pj}(E )Z_{j\ell}(E)\over 1-Z_{pp}(E) }
{\ln (\Lambda_j/\Lambda_p) \over 1-\Lambda_p/\Lambda_j }
{ m_j\, c\, h_0\over E\, \tau_j}\ f(\theta) \phi_p(E,0) ,
\end{equation}
\begin{equation}
\phi_\ell^{j,low} =  {Z_{pj}(E)
Z_{j\ell}(E)
\over 1-Z_{pp}(E)} \phi_p(E,0)\ ,
\end{equation}
where an isothermal model for the atmosphere, in which
$\rho (h)=\rho_0 \exp(-h/h_0)$ describes the density
profile as a function of altitude $h$. The parameters are
$h_0=6.4$ km and $\rho_0=2.03\times 10^{-3}$ g/cm$^3$ \cite{atm}.
The quantity $m_j$ is the decaying particle's mass and
\begin{equation}
\Lambda_j\equiv {\lambda_j\over (1-Z_{jj})}
\end{equation}
is an effective interaction length, which is weakly dependent on energy.
The zenith
angle dependence of the high energy
flux is characterized by $f(\theta)\simeq 1/\cos\theta$ for $\theta<60
^\circ$.
At higher zenith angles, $f(\theta)$ is a more complicated function which
accounts for the curvature of the earth. Details appear in Ref. \cite{lipari}.
The low energy flux is isotropic.
When the cascade involves charmed hadrons, the low energy behavior
dominates and the flux is called ``prompt''. Critical energies, below
which the decay length is less than the vertical depth of the atmosphere, range
from $3.7-9.5\times 10^{7}$ GeV \cite{tig}.
Interpolation between high and low energy fluxes is done via
\begin{equation}
\phi_\ell=\sum_j {\phi_\ell^{
j,{low}}\phi_\ell^{j,{high}}\over
\phi_\ell^{j,{low}}+\phi_\ell^{j,{high}}}\ .
\end{equation}

Eqs. (2.6) and (2.7) show that the bases for the calculation of the
prompt lepton fluxes are production and decay $Z$-moments involving charm.
The decay moments are discussed in Section III.D. 
The main uncertainties in the calculation of the lepton flux from atmospheric
charm are the production $Z$-moments: $Z_{pD}$ and  $Z_{p\Lambda_c}$.
The production moments
are given by
\begin{equation}
Z_{pc}=2\int_0^1{dx_E\over x_E}{\phi_p(E/x_E)\over\phi_p(E)}{1\over
\sigma_{pA}(E)}{d\sigma_{pA\rightarrow c\bar{c}}(E/x_E)\over
dx_E}\ .
\end{equation}
The differential cross section is evaluated here using perturbative QCD.
The factor of two accounts for the multiplicity of charmed (or anticharmed)
particles. 
The charm $Z$-moments can be converted to hadronic moments by
\begin{equation}
Z_{pj}(E)=f_j \, Z_{pc}(E)\ ,
\end{equation}
where $f_j$ is the fraction of charmed particles which emerges as
hadron $j$, where $j=D^0,\ D^+,\ D_s^+$ and $\Lambda_c$. We 
implicitly sum over 
particles and antiparticles (hence the factor of two in Eq. (2.10)).

The inelastic proton-air cross section $\sigma_{pA}(E)$ is parameterized
by \cite{mielke}
\begin{equation}
\sigma_{pA}(E)= 280 - 8.7\ln(E/{\rm GeV})
+1.14 \ln^2(E/{\rm GeV})\ {\rm mb}\ .
\end{equation}
In the high energy limit of the lepton fluxes, in addition to
$Z_{pc}$, we need effective hadronic interaction lengths $\Lambda_j$.
The proton effective interaction length is therefore
\begin{equation}
\Lambda_p(E)\simeq {A\over N_0\sigma_{pA}(E)}{1\over (1-Z_{pp})}
\end{equation}
where $A=14.5$ is the average atomic number of air nuclei
and $N_0=6.022\times 10^{23}$/g.
We use Thunman {\it et al.}'s (TIG) energy dependent 
$Z_{pp}$, calculated using a PYTHIA Monte Carlo \cite{pythia} 
as a function of energy \cite{tig}.
The charmed hadron $j$ interaction lengths are all taken to be equal to the
kaon interaction length, approximated by
\begin{equation}
\Lambda_j\simeq {A\over N_0\sigma_{pA}(E)}{\sigma_{pp}^{tot}(E)\over
\sigma_{Kp}^{tot}(E)}{1\over(1- Z_{KK})}\ .
\end{equation}
We use $Z_{KK}$ from Ref. \cite{tig}.
The total cross sections are parameterized using the particle data
book values \cite{pdg} based on Regge theory \cite{dl}.
The prompt lepton flux below $10^8$ GeV is insensitive to the detailed
values of $\Lambda_j$ because essentially all of the charmed hadrons decay
before reaching the surface of the earth. 
Therefore, for most of the energy range considered here, the charmed
particles are ``low energy'' and Eq. (2.7) describes the lepton fluxes.

We now turn to the evaluation of $Z_{pc}$ in perturbative QCD and
the other charm inputs.

\section{Charm Cross Section and Energy Distribution}

The charm production cross section and energy distribution are
the largest uncertainties in the calculation of the prompt lepton fluxes.
Since the charm quark mass is of the order of $1.3$ GeV, 
the treatment of the charm quark as a heavy quark may be 
questionable.  
Theoretical uncertainties,
due to the possible range of charm quark masses,
as well as the usual 
factorization and renormalization scale dependence need to be 
studied.  Theoretical
predictions based on perturbative QCD calculation 
fit the available data reasonably well 
in the energy range up to 800 GeV beam energy \cite{data}.  
However, 
atmospheric lepton flux calculations require
beam energies up to and beyond $10^8$ GeV. The
parton distribution functions are needed 
at very small parton momentum fraction $x$, outside of the
measured regime \cite{hera}.

In this section we will address these theoretical issues:
\begin{itemize}
\item the effect of next-to-leading order corrections
on the cross section and charmed particle energy distribution,
\item charmed quark mass dependence,
\item factorization and renormalization scale dependence,
\item the consequences of the small-$x$ behavior of the parton distribution
functions on the interaction $Z_{pc}$ moment, and
\item the $A$ dependence of the proton-air charm production
cross section.
\end{itemize}
From our evaluation of these quantities, a theoretical uncertainty
associated with perturbative charm production will be evaluated.
We also describe our inputs to the decay moments of charmed hadrons.

\subsection{Total Cross Section}

The next-to-leading order (NLO) total charm cross section has been calculated
by Nason, Dawson and Ellis \cite{nde} and by van Neerven
and collaborators \cite{smith}.
The NLO cross section is a factor of between 2 and 2.5 larger than the leading
order cross section.
Gluon fusion dominates the production process.
In Fig. 1, we show the importance of the charm quark mass in the
NLO cross section.  We compare the NLO 
$\sigma(pN\rightarrow c\bar{c}X)$ 
as a function of the beam energy $E$ 
obtained with the renormalization scale $\mu$ equal to 
the factorization scale $M$ equal to the charm quark mass $m_c$ with
$m_c=1.3$ GeV and $m_c=1.5$ GeV.   The cross sections are evaluated
using the CTEQ3 parton distribution functions \cite{lai}.
The corresponding value of $\Lambda_4^{\overline{MS}}$ is 239 MeV.
Fixed target data from a summary by Frixione {\it et al.} \cite{fmnr}
are also plotted.
We note 
that the fixed target data seem to prefer $m_c=1.3$ GeV. 
In all of the subsequent figures, we set $m_c=1.3$ GeV. 
The CTEQ3 parton distribution functions
will be our canonical set, in part because they incorporate global fits
to HERA data, and while their validity is not claimed for
parton fraction $x$ below $x_{min}=10^{-5}$ and $Q_0=1.6$ GeV,
the program nevertheless provides
smooth parton distribution functions below these values.

In Fig. 2 we show dependence of the total cross section on the 
scale
and parton distribution.  
We plot the NLO cross section
for different values of $\mu$ and $M$:
using the
CTEQ3 structure functions, we set 
$\mu=M= m_c$ (dot-dashed)
and $\mu=m_c$, $M=2 m_c$ (solid) with $m_c=1.3$ GeV. The 
dashed line is the cross section obtained with the
MRSD- parton distribution functions \cite{mrsdm} and scales $\mu=m_c$,
$M=2 m_c$ with $m_c=1.3$ GeV. Also plotted are the data as in Fig. 1.

The MRSD- distribution functions have
a small-$x$ behavior that is suggested by the BFKL approach
\cite{bfkl}. In the
small-$x$ limit,
the parameterization of the gluon (and sea quark) distribution
functions at reference scale $Q_0$ is
\begin{equation}
xg(x,Q_0)\sim x^{-\lambda}\ .
\end{equation}
The D- distributions have  $\lambda = 0.5$.
Typically, global fits such as
the MRSA \cite{mrsa}, MRSG\cite{mrsg} and
CTEQ3 distributions have $\lambda\simeq 0.3$.
By using the D- distributions, we are effectively setting an upper limit
on the perturbative charm cross section, given our choices of $m_c$, $\mu$ and
$M$. We note that, generally, parton distribution functions begin
evolution at $Q_0$ larger than 1.3 GeV. Consequently, our default factorization
scale is $M=2m_c$ so that we can use more than the CTEQ3 parameterizations.

Fig. 2 indicates that a low energies,
the total cross section has weak dependence on 
the choice of the scale and the parton distribution function. 
At high energies, 
$E\geq 10^6$ GeV, there is a factor of 1.7-2.1 increase from
$M=m_c$ to $M=2m_c$. The D- cross section is a factor of 1.3
larger than the CTEQ3 cross section at $E=10^6$ GeV, both with
$M=2m_c$. The D- cross section increases more rapidly because of
the steeper small-$x$ behavior of the parton distribution function and is
enhanced by a factor of 2.6 at $10^8$ GeV. This gives an
overall uncertainty of factor 
of 5.5 at the highest energy of $10^8$ GeV.  The MRSA and MRSG
cross sections for $M=2m_c$ lie between the upper and lower curves
in Fig. 2.

The total charm cross section in p-Air collisions, 
$\sigma_{pA\rightarrow c\bar{c}} (E)$, can be written as 
\begin{equation}
\sigma_{p A\rightarrow c\bar{c}} = A^\gamma \sigma_{pN\rightarrow c\bar{c}}
\end{equation}
We have evaluated the $A$ dependence for charm pair
production using a Glauber-Gribov model of nuclear shadowing 
\cite{valerio}.  
We find that over an energy range of $10^2-10^6$ GeV,
$\gamma=1.0-0.8$. Since $A=14.5$, the shadowing effect is small,
so we set $\gamma=1$. This is consistent with recent measurements at $E=800$ GeV
\cite{gamcc}. Low energy measurements at larger $x_E$
\cite{gamc} indicate smaller $\gamma$ values ($\gamma\simeq 0.75$),
which would reduce our flux predictions by an overall factor of 0.5.

We have used a comparison between data and theory for
the total cross section to show that $m_c=1.3$ GeV is a
reasonable choice, and to estimate the range of cross sections, related to
the approximate uncertainty in the flux. To evaluate $Z_{pc}$, we need the
energy distribution of the charmed particle. In the next section, we discuss
the
energy distribution of charm quarks in NLO QCD.

\subsection{Charm Energy Distribution}

NLO single differential distributions 
in charm quark production have been evaluated Nason, Dawson and
Ellis \cite{nde2} and incorporated into a computer program, which
also calculates double differential distributions,
by Mangano, Nason and Ridolfi \cite{mnr}. The program is time
consuming, so we have incorporated NLO corrections to
$d\sigma/dx_E$ by rescaling the leading order distribution.
The $x_E$ distributions at next-to-leading order are well fit by
a $K$-factor rescaling which is a function of $x_E$, where $K$ is
defined by
\begin{equation}
K\equiv {d\sigma(NLO)/dx_E\over d\sigma(''LO'')/dx_E}
\end{equation}
where ``$LO$'' means taking the leading order matrix element squared,
but using the two-loop $\alpha_s(\mu^2)$ and the NLO parton distribution
functions. 
$K$ defined this way shows the effects of the NLO matrix element corrections.

Using the NLO computer program with
the CTEQ3 parton distribution functions,
we show our results for $K(E,x_E)$ for 
$E=10^3$ and $2K(E,x_E)$ for $E=10^6$ GeV in Fig. 3.  
We find that $K$ evaluated using the D- and MRSA distributions
agree well with Fig. 3.
The error bars indicate the
numerical errors associated with the Monte Carlo integration in the
NLO program. At higher energies, the errors become larger
for comparable $x_E$ because the cross section is dominated by small
$x_E$. $K$ can be parameterized as 
\begin{eqnarray}
K(E,x_E) & = & 1.36+0.42\ln(\ln(E/{\rm GeV}))\\ \nonumber
& &\quad +\Bigl( 3.40+
18.7(E/{\rm GeV})^{-0.43}-0.079\ln(E/{\rm GeV})\Bigr)\cdot x_E^{1.5}
\end{eqnarray}
for $\mu=m_c$ and $M=2m_c$.
The parameterization is shown by the curves in the figure.

Using the $x_E$ and energy dependent $K$-factor,
we plot the charm quark $x_E$ distribution for $E=10^3$ GeV, $10^6$ GeV
and $10^9$ GeV in Fig. 4.
The distributions fall rapidly with $x_E$. The convolution of the
differential distribution with the ratio of proton fluxes and interaction lengths, integrated
over $x_E$ at fixed outgoing charm quark energy, is what is
required for the $Z$-moment.

Fig. 5 shows the differential moment,
$dZ_{pc}/dx_E$. The solid lines are for CTEQ3 distributions at
outgoing charm energies $E=10^4$ GeV, $10^6$ GeV and $10^8$ GeV,
in increasing magnitudes. The dashed lines represent the same quantities
for the D- calculation.
The D- distributions have approximately the same shape as the 
CTEQ3 distributions, but there is a more rapid growth in overall
normalization with
energy.

In the context of 
perturbative hard scattering production of charm pairs, 
the average $x_E$ value in the evaluation of $Z_{pc}$ is 0.15-0.2.
More than 80\% of the cross section comes from charm transverse momenta
below a value of $2m_c$. In the low transverse momentum limit, $x_E\simeq x_F$.
Fixed target experiments measure $d\sigma/dx_F$. The measured
charmed meson $x_F$
distributions are consistent with the perturbative NLO QCD calculations
for charm quark production, without any fragmentation corrections that
would soften the $x_F$ distributions \cite{xf}. Fragmentation calculations
are applicable at large transverse momentum. For the calculation of 
$Z_{pc}$, we are in the low transverse momentum regime, so we do not
need fragmentation. 

\subsection{Hadron Fractions}

We account for the transformation of charmed
quarks into hadrons by an energy independent hadronic fractions.
The hadronic fractions convert $Z_{pc}$ into the interaction moments
for the charmed mesons and the $\Lambda_c$ via Eq. (2.11).
The hadron fractions can be obtained by the observation that \cite{fmnr,aoki}
\begin{eqnarray}
\sigma (D_s)&\simeq & 0.2\ \sigma(D^0+D^+)\\
\sigma(\Lambda_c) &\simeq & 0.3\ \sigma(D^0+D^+)\ .
\end{eqnarray}
The fractions of charmed quarks that appear as $D_s$ and $\Lambda_c$ are
\begin{eqnarray}
f_{D_s} &=& 0.13\ ,\\
f_{\Lambda_c} &=& 0.20\ . 
\end{eqnarray}
To get the $D^+$ and $D^0$ fractions, 
one needs a ratio of $\sigma(D^+)/\sigma(D^0)$.
Using arguments
based on isospin invariance and counting of states in the
production of $D$ and $D^*$, together with branching fractions for
$D^*\rightarrow D$, Frixione {\it et al.} \cite{fmnr} suggest that
$\sigma(D^+)/\sigma(D^0)\simeq 0.32$. With this assumption for the
ratio of the cross sections,
\begin{eqnarray}
f_{D^0} &=& 0.51\ ,\\
f_{D^+} &=& 0.16\ . 
\end{eqnarray}

There is some uncertainty in the values of $f_j$.
Experimental measurements of $\sigma(D^+)/\sigma(D^0)$ in $pN$ and
$pp$ fixed target experiments tend to lead to a somewhat higher
ratio of cross sections. For example in $pp$ collisions with
a beam energy of $E_b=400$ GeV, the LEBC-EHS Collaboration \cite{lebc} 
measures $\sigma(D^+)/\sigma(D^0)=0.7\pm 0.1$, while in $pN$ collisions
at $E_b=250$ GeV, the ratio is measured by E769 \cite{esss}
to be $0.57\pm 0.22$.
By taking $\sigma(D^+)/\sigma (D^0)=0.6$, 
the resulting change in the predicted flux is only $\sim 15\%$.

Integrated $Z$-moment
$Z_{pD^0}$ scaled by $10^3$  versus charmed particle
energy is shown in Fig. 6
for the D- and CTEQ3 distributions with
$\mu=m_c$ and $M=2 m_c$.  Also shown is 
the CTEQ3 calculation with $\mu=M=m_c$. The other $Z$-moments for
charm production are simple rescalings of Fig. 6. While the curves are
similar up to 1 TeV, by $E=10^{6}$ GeV, there is a factor of $\sim 5$ between
the upper and lower curves. At $E=10^8$ GeV, the upper and lower curves
differ by more than a factor of 10, larger than the ratio of
cross sections at the same energy. This is accounted for by the fact that
the $Z$-moments at energy $E$ involve integrals of the cross section at
a higher energy. In addition, since $x_E\sim 0.15-0.2$ and the cross section
is dominated by parton invariant masses near $m_c$, small parton $x$ values are
emphasized. Since the 
prompt flux is proportional to $Z_{pc}$, this enhancement
is reflected in the flux as well.

\subsection{Decay Moments}

The last elements of the calculation of
the lepton fluxes from charm are the decay
moments $Z_{kl}(E)$ for $k=D^+,\ D^0,\ D_s^+$ and $\Lambda_c$.
The decay moments can be written in the same form as Eq. (2.4) with 
$\lambda_k$ now representing the decay length. The decay distribution
can be represented by 
\begin{equation}
{dn_{k\rightarrow \ell}(E;E_k)\over dE}={1\over E_k}F_{k\rightarrow \ell}
\Biggl( {E\over E_k} \Biggr)\ ,
\end{equation}
so the decay moments, in terms of an integral over $x_E=E/E_k$, are
\begin{equation}
Z_{k\ell}(E) = \int dx_E {\phi_k(E/x_E)\over \phi_k(E)} F_{k\rightarrow \ell}
(x_E)\ .
\end{equation}
The function $F$ is given in Ref. \cite{lipari,bugaev}, in the approximation 
that the
leptonic decays of charmed mesons are approximated by three-body
decays. 
In the ultrarelativistic limit, $F$ is nearly equal for
$\ell=\nu_e\ ,\nu_\mu$ and $\mu$, so the decay moments for the three
leptons are essentially equal. Consequently, 
we take the fluxes for the three leptons
to be equal.

Following Bugaev {\it et al.} in Ref. \cite{bugaev}, the effective hadronic
invariant mass $m_X$ of the decay of the $D^+$ is taken to be
630 MeV and for the $D^0$ decays, 670 MeV. We take $m_X=840$ MeV for
$D_s$ decays. 
The $\Lambda_c$ will ultimately contribute little over most of the energy
range considered. For $\Lambda_c$ decays, we use the same three-body 
formula with an effective hadronic mass of 1.3 GeV.
The branching ratios are:
\begin{eqnarray}
B(D^+\rightarrow \ell ) &=& 17\% \ ,\\ \nonumber
B(D^0\rightarrow \ell ) &=& 6.8\% \ ,\\ \nonumber
B(D_s\rightarrow \ell ) &=& 5.2\% \ ,\\ \nonumber
B(\Lambda_c\rightarrow \ell ) &=& 4.5\% \ .
\end{eqnarray}

For the energies considered here, 
the charmed particle fluxes in the low energy limit
dominate.  
Assuming that $Z_{pp}$ and $\Lambda_p$ are nearly energy independent,
this means that
\begin{equation}
\phi_k\sim Z_{pk}(E)\, E\phi_p(E)\ .
\end{equation}
The proton flux falls like $E^{-2.7}-E^{-3}$. The charm
production $Z$-moments increase with energy, as seen in Fig. 6.
When we put in the low energy $D^+$ meson flux and evaluate the
$Z_{D^+\ell}$ moment, we get the results shown in Fig. 7. All of the
other low energy
decay moments can be obtained by branching fraction rescaling.
For the high energy moments, we take $Z_{pk}\sim E^{0.42}$ for the D-
distributions and $Z_{pk}\sim E^{0.23}$ for CTEQ3,
with $\phi_k\sim Z_{pk}(E)\phi_p(E)$.

\section{Prompt Lepton Flux}

In Fig. 8 we show our results for the prompt atmospheric flux
scaled by $E^3$ for two parton distributions and factorization
scale choices.  The highest flux  at
$E=10^8$ GeV is with the D- distribution and
$M=2\mu=2 m_c$ (dashed). The CTEQ3 distributions with the same choice
of scale are represented by the solid line, while the dot-dashed line
shows the result when $M=\mu=m_c$. For reference, we show the vertical
conventional and prompt flux calculated and parameterized by TIG in
Ref. \cite{tig}. The fluxes directly reflect the interaction
$Z$-moments of Fig. 6. We emphasize that the prompt flux is isotropic
except at the highest energies, while the conventional flux is not.

We have also estimated the flux due to pion-air interactions creating
charm pairs. The effect is to increase the prompt flux by $\sim 30\%$
at $10^2$ GeV and by $\sim 15\%$ at $10^6$ GeV. This is a small
effect, so we neglect pion contributions to charm production.

The prompt lepton flux evaluated using perturbative QCD can be
parameterized as 
\begin{equation}
\log_{10}\Bigl( E^3\phi_\ell (E)/({\rm GeV}^2/
{\rm cm}^2\,{\rm s\, sr})\Bigr)
 = -A + B\, x+C\,x^2-D\, x^3\,
\end{equation}
where $x\equiv \log_{10}(E/{\rm GeV})$.
In Table I, we collect the constants for the D- and CTEQ3 fluxes
exhibited in Fig. 8.

The TIG flux relies on PYTHIA calculations with $m_c=1.35$ GeV and
the MRSG \cite{mrsg} parton distribution functions.
For $E<1$ TeV, the TIG flux is larger than ours because of fragmentation
effects. Their calculation uses the Lund hadronization model \cite{lund}
which can give the charmed hadron a larger energy than the charmed
quark, an effect which is larger for smaller center of mass energies.
The TIG prompt flux calculation is lower than our calculation for
energies $E> 1$ TeV. 
At parton $x<10^{-4}$, the parton distributions
in the TIG calculation are flattened. For example,
the gluon distribution is
\begin{equation}
xg(x,Q)\sim x^{-0.08}
\end{equation}
below $x=10^{-4}$.
We note that HERA data at $10^{-5}<x$ and $Q_0\sim 2$ GeV show no indication
of flattening \cite{hera}.
NLO effects in the TIG calculation are accounted for by an
energy and $x_E$ independent factor of 2. Overall, the net effect is that
the TIG $Z_{pc}$ is nearly energy independent.

The prompt muon flux by Volkova {\it et al.} \cite{vfgs} is larger than
our calculated flux. This comes in part because of their assumption
that $d\sigma/dx_E\sim (1-x_E)^5/x_E$, independent of center of mass 
energy and a cross section larger than the perturbative one below
$E\sim 10^5$ GeV. This $x_E$ dependence is harder than the
perturbative $x_E$ dependence shown in Fig. 4.

Bugaev {\it et al.} \cite{bugaev} have presented calculations of the prompt
muon flux using two phenomenological, nonperturbative approaches.
One is based on the Recombination Quark-Parton Model (RQPM) and
the other on the Quark-Gluon String Model (QGSM). The QGSM prompt flux
is relatively small compared to the RQPM flux, which already affects the
total atmospheric muon flux at energies of a few tens of TeV.
Relative to the RQPM calculation  and the
Volkova {\it et al.} results, our D- prompt flux is lower.

Several experiments show an excess in muon flux above $\sim 10$ TeV
\cite{rhode,kgf,other}. Following Rhode in Ref. \cite{rhode}, we plot in
Fig. 9 the quantity
$E^{3.65} \phi_\mu(E)$, where $\phi_\mu$ represents the sum of
the prompt and vertical conventional flux. Also shown are the data 
from Ref. \cite{rhode}.
The energy scale factor
mostly
accounts for the rapidly falling conventional flux \cite{comment}.
When we add the
prompt fluxes of Fig. 8 to the TIG vertical conventional flux, one
sees an enhancement at muon energies above $10^5$ GeV, at a higher
energy than the experimental excess shown by data points. 

In Ref. \cite{prs}, we have shown that it is possible to
enhance the prompt flux sufficiently to account for some of the
observed muon excess at a few TeV. This is accomplished by
extrapolating the charm cross section at 1 TeV with a
faster growth in energy than predicted by perturbative QCD.
The $x_E$ dependence was taken as
$d\sigma/dx_E\sim
(1-x_E)^4$.
The inputs are consistent with fixed target data
below 1 TeV beam energies. We found that the
predicted prompt flux made significant contributions
in the region of the observed excess of muons, but it does not
fully describe the Fr\'ejus data \cite{rhode}.
These inputs are not consistent with perturbative QCD. The
experimental excess of muons
cannot be accounted for by perturbative QCD production of charm.

Another implication  of the prompt fluxes calculated here is the ratio
of the muon neutrino to electron neutrino flux. We
define
\begin{equation}
R_\theta ={\phi_{\nu_\mu}\over \phi_{\nu_e}}\ .
\end{equation}
Using the TIG parameterization of the vertical $\theta=0^\circ$ neutrino
fluxes, adding conventional and prompt contributions, $R_0$ 
as a function of energy is shown in Fig.
10. At a zenith angle of $60^\circ$, the conventional flux is a factor
of two larger. The ratio $R_{60}$ is also shown in Fig. 10.
The quantity $R$ is an interesting diagnostic for the onset of
prompt neutrino dominance. Unfortunately, at these energies,
$R$ is difficult to measure.

\section{Conclusions}

We find that the perturbative charm contributions to lepton
fluxes are significantly larger than the recent TIG calculation.
The prompt muon flux becomes larger than the conventional muon
flux from pion and kaon decays at energies above $\sim 10^5$ GeV.
We set values of the charmed quark mass, renormalization scale
and factorization scale by fitting the
charm production cross section to low-energy data, then we extrapolate
to higher energies. We find that the NLO corrections give a correction
of more than a factor of two which is weakly energy and $x_E$
dependent. Nuclear shadowing corrections are small for all
energies, due to the air nucleus being relatively light.
The main uncertainty in the perturbative calculation of the
prompt flux, given fixed charm mass, factorization scale and
renormalization scale, is the small-$x$ behavior of the parton
distribution functions. Different choices of scales and
distribution functions, extrapolated to low $x$ with the
same power law dependence as for $x>10^{-5}$,
yield as much as a factor of $\sim 10$ discrepancy in
the prompt flux at $E=10^8$ GeV. 

If the parton distributions are flattened
below some critical value $x<x_c$ according to Eq. (4.2), our curves
in Fig. 8 would overestimate the lepton fluxes. For example, if
$x_c=10^{-5}$, the calculated prompt flux at
$E=10^6$ GeV, using CTEQ3 with $M=2\mu=2m_c$, is $\sim 80$\% of the
value shown by the solid curve in Fig. 8, reducing to $\sim 40$\%
of the value at $10^8$ GeV. For $x_c=10^{-6}$, the flux is $\sim 70$\%
of our calculated value at $10^8$ GeV. However,
such an abrupt turnover in the
power law behavior of the small-$x$ parton distributions is unlikely.

We conclude that the prompt muon flux calculated in the context
of perturbative QCD cannot explain the observed excess of
muons in the TeV region \cite{rhode,kgf,other}, independent
of the theoretical uncertainties associated with small
parton $x$. However, prompt fluxes calculated using
non-perturbative models of charm production such as discussed in
Refs. \cite{bugaev,prs} could provide a muon excess in that energy
range. Measurements of the atmospheric flux in the 100 TeV
range would help pin down the charm cross section at energies
above those currently accessible using accelerators and would
provide valuable information about the small-$x$ behavior of
the gluon distribution function.

Even though the prompt contributions to the lepton fluxes
change the energy behavior of the differential fluxes by
a factor of $E$, the atmospheric neutrino fluxes do not
compete with neutrino fluxes from extragalactic sources
above 10 TeV \cite{gqrs}.
Possible oscillations of muon neutrinos as indicated by
the Super-Kamiokande experiment \cite{superk} do not affect
our results due to the extremely small oscillation probability
for the energies of interest.

\acknowledgements
Work supported in part
by National Science Foundation Grant No.
PHY-9507688 and D.O.E. Contract No. 
DE-FG02-95ER40906. M.H.R. acknowledges the hospitality of the CERN
Theory Division where part of this work was completed. We thank 
P. Gondolo, P. Nason, W. Rhode, D. Seckel and
M. Mangano for useful discussions, the Aspen Center for Physics
and M. Mangano for providing a copy
of the Mangano, Nason and Ridolfi NLO computer program.

% TABLE

\begin{table}
\caption{Parameters for the prompt
muon plus antimuon flux appearing in Fig. 8:
$\log_{10}(E^3\phi_\mu/({\rm GeV}^2/{\rm cm}^2\,{\rm s\ sr}))=
-A+B\, x+C\,x^2-D\,x^3$, where $x\equiv \log_{10}(E/{\rm GeV})$.}
\begin{tabular}{llcccc}

PDF & Scales & $A$ & $B$ & $C$ & $D$ \\ \hline
CTEQ3 & $M=\mu=m_c$ & 5.37 & 0.0191 & 0.156 & 0.0153 \\
CTEQ3 & $M=2\mu=2m_c$ & 5.79 & 0.345 & 0.105 & 0.0127 \\
D- & $M=2\mu=2m_c$ & 5.91 & 0.290 & 0.143 & 0.0147 \\
\end{tabular}
\end{table}

%figure 1
\begin{figure}
\centerline{\psfig{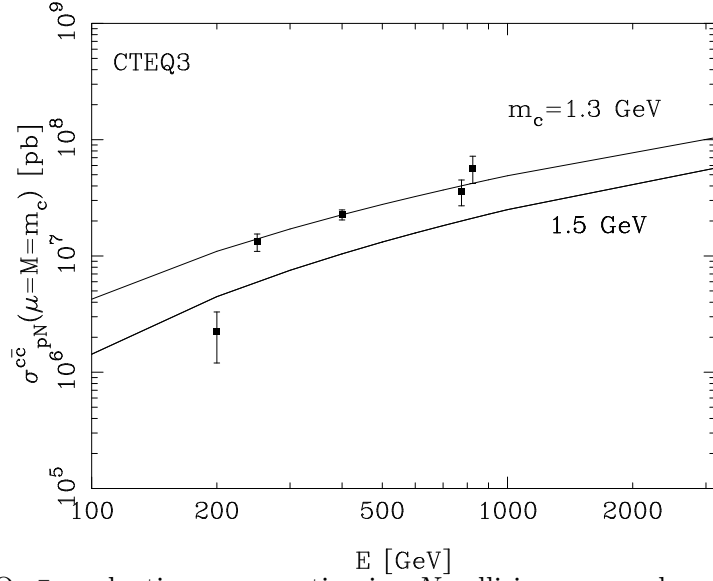}}
\caption{The NLO 
$c\bar{c}$ production cross section in $pN$ collisions
versus beam energy for $m_c=1.3$ and 1.5 GeV. The CTEQ3 parton 
distribution functions are used with $M=\mu=m_c$.
The data are taken
from the summary in Ref. [28].}
\end{figure}

%figure 2
\begin{figure}
\centerline{\psfig{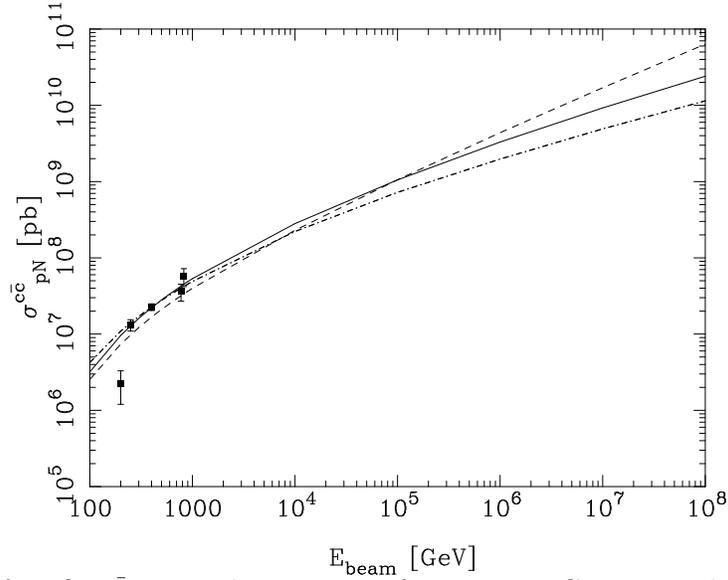}}
\caption{A plot of  NLO $\sigma _{pN}^{c\bar{c}}$ versus 
beam energy for $m_c=1.3$ GeV using the CTEQ3 (solid) and D- (dashed)
parton distribution functions
with $M=2m_c$ and $\mu=m_c$. Also show is the CTEQ3 NLO prediction
with $M=\mu=m_c$ (dot-dashed). The data are the same that appear in Fig. 1.}
\end{figure}
\vfil\eject

%figure 3
\begin{figure}
\centerline{\psfig{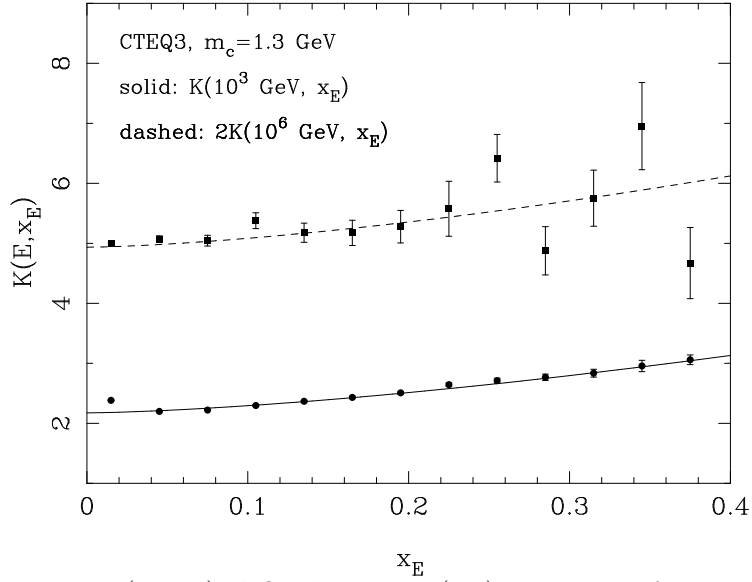}}
\caption{The function $K(E,x_E)$ defined in Eq. (3.3) versus $x_E$ for
$E=10^3$ GeV and $10^6$ GeV. The points come from the evaluation of $K$
using the results of Ref. [41,42] with error bars indicating numerical
errors in the integration, and the curves are our fit to the ratio 
parameterized in Eq. (3.4).
}
\end{figure}

%figure 4
\begin{figure}
\centerline{\psfig{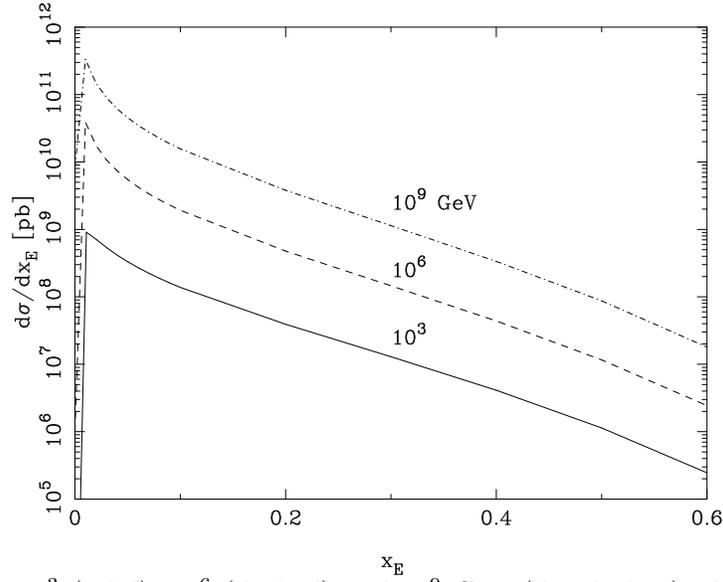}}
\caption{For $E=10^3$ (solid), 10$^6$ (dashed) and 10$^9$ GeV
(dot-dashed), $d\sigma/dx_E$, including
the factor of $K(E,x_E)$. The scales used are $\mu=m_c$ and $M=2m_c$,
for $m_c$=1.3 GeV.
}
\end{figure}

\vfil\eject

%figure 5
\begin{figure}
\centerline{\psfig{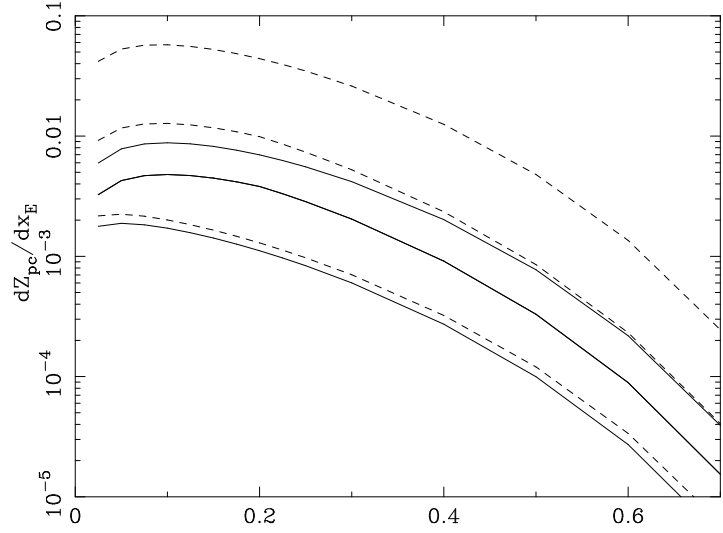}}
\caption{
For energies
$E=10^4$ GeV, $10^6$ GeV and $10^8$ GeV, 
$dZ_{pc}(E)/dx_E$ versus $x_E$ for CTEQ3 (solid) and D- (dashed)
parton distribution functions, where $\mu=m_c$ and $M=2m_c$.
}
\end{figure}

%figure 6
\begin{figure}
\centerline{\psfig{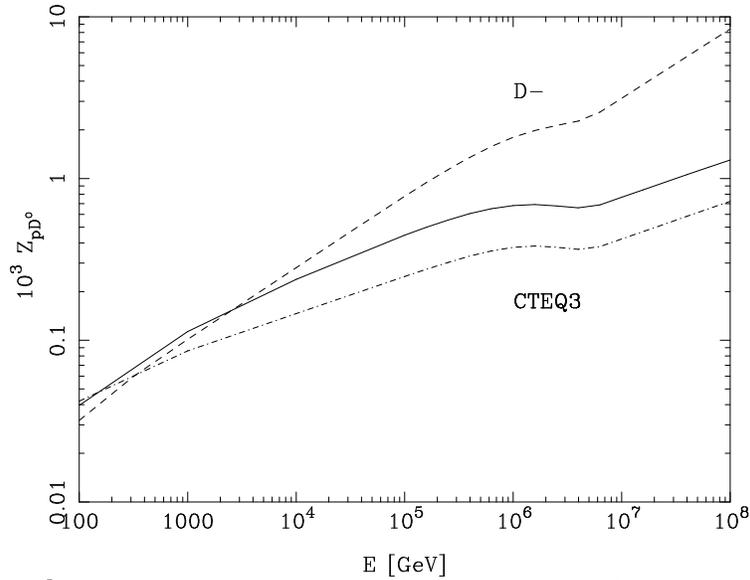}}
\caption{
$Z_{pD^0}\times
10^3$ versus $E$ for CTEQ3 (solid) and D- (dashed) parton distribution
functions with $\mu=m_c$ and $M=2m_c$. Also shown is $Z_{pD^0}\times 10^3$
for CTEQ3 with $\mu=M=m_c$ (dot-dashed).
}
\end{figure}

\vfil\eject

%figure 6prime
\begin{figure}
\centerline{\psfig{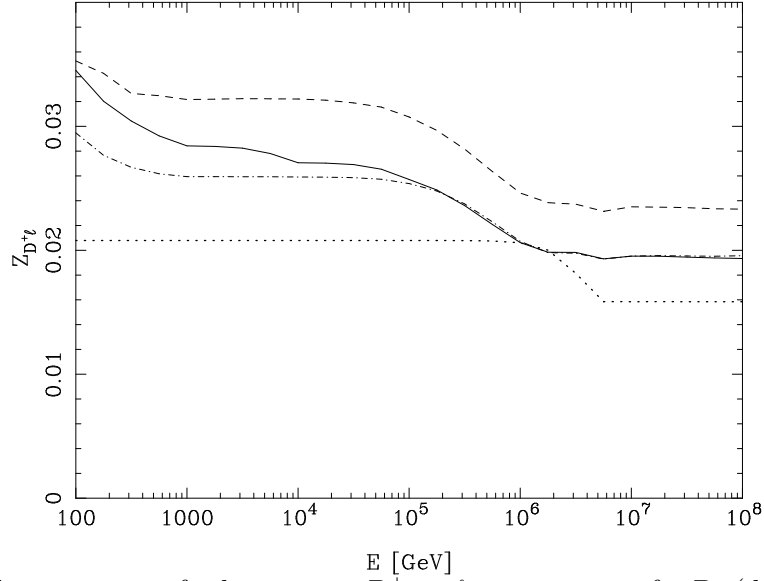}}
\caption{
The decay moment for low energy $D^+\rightarrow \ell$ versus energy for
D- (dashed) and CTEQ3 (solid) with $M=2\mu=2m_c$ and for
CTEQ3 with $M=\mu=m_c$ (dot-dashed). The dotted line indicates the
decay moment if $Z_{pc}$ is taken independent of energy.
}
\end{figure}

%figure 7
\begin{figure}
\centerline{\psfig{figure=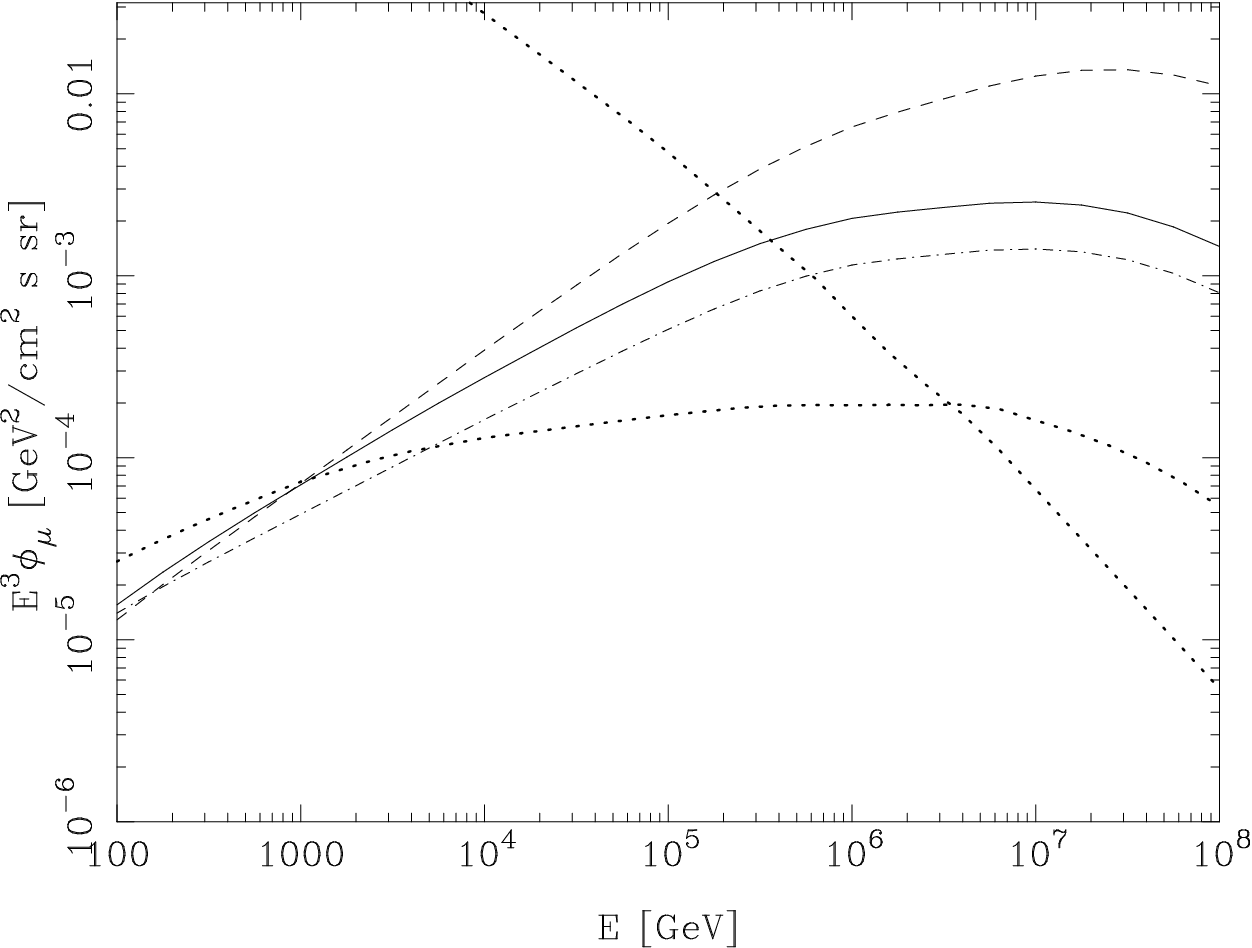,height=3.0in,angle=270}}
\caption{
The prompt atmospheric muon flux scaled by $E^3$ versus muon energy
for CTEQ3 (solid) and D- (dashed) with $M=2\mu=2m_c$. Also shown is the
scaled muon flux using CTEQ3 with $M=\mu=m_c$ (dot-dashed) and the
TIG parameterization of the prompt muon flux and the vertical
conventional muon flux (dotted).
}
\end{figure}

%figure 9
\begin{figure}
\centerline{\psfig{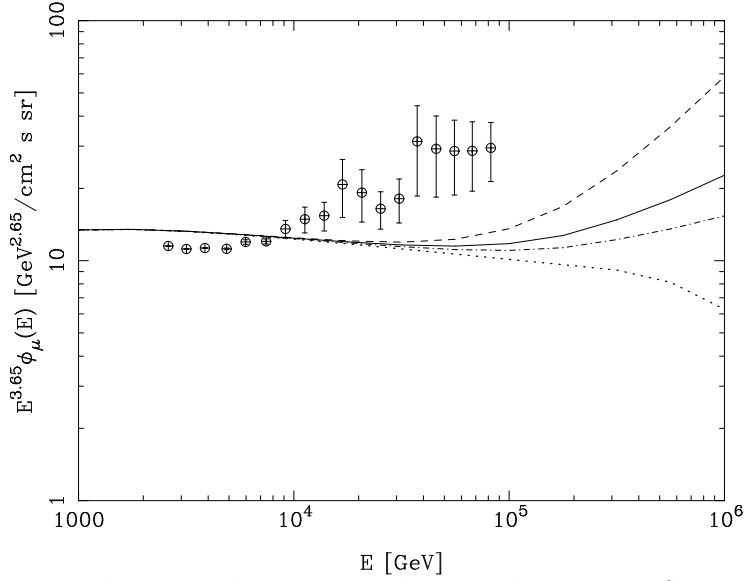}}
\caption{
The prompt plus vertical conventional
atmospheric muon flux scaled by $E^{3.65}$ versus muon energy
for CTEQ3 (solid) and D- (dashed) with $M=2\mu=2m_c$. Also shown is the
scaled muon flux using CTEQ3 with $M=\mu=m_c$ (dot-dashed) and the
TIG parameterization of vertical conventional muon flux 
(dotted). The data shown are from Ref. [7].
}
\end{figure}

%figure 10
\begin{figure}
\centerline{\psfig{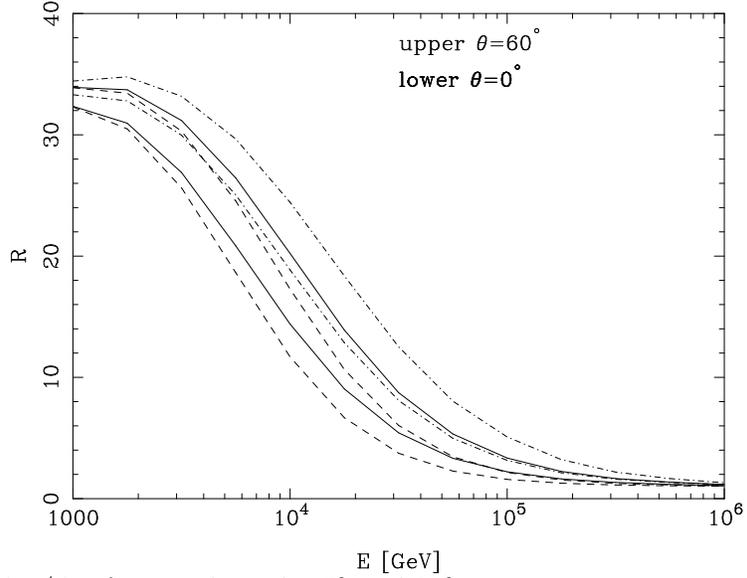}}
\caption{
$R=\phi_{\nu_\mu}/\phi_{\nu_e}$ for zenith angles $0^\circ$ and
$60^\circ$ versus neutrino energy
for CTEQ3 (solid) and D- (dashed) with $M=2\mu=2m_c$, and
using CTEQ3 with $M=\mu=m_c$ (dot-dashed). 
}
\end{figure}


\begin{references}
\bibitem{comment}
A better choice of rescaling given our assumed cosmic ray flux would be
$E^{3.7}\phi_\mu$.
\end{references}
\end{document}